\documentclass[final,5p,twocolumn,times,sort&compress]{elsarticle}

\def\preprintdate{IUHET 622, November 2016}

\def\tfrac#1#2{{\textstyle{{#1}\over {#2}}}}
\def\notag{\nonumber}
\def\widehat{\hat}
\def\text#1{{\rm #1}}

\def\al{\alpha}
\def\be{\beta}
\def\ga{\gamma}
\def\de{\delta}
\def\ep{\epsilon}

\def\et{\eta}
\def\th{\theta}

\def\ka{\kappa}
\def\la{\lambda}

\def\rh{\rho}

\def\si{\sigma}

\def\ph{\phi}
\def\vp{\varphi}
\def\ch{\chi}

\def\om{\omega}

\def\La{\Lambda}

\def\cL{{\cal L}}

\def\half{{\textstyle{1\over 2}}}
\def\quar{{\textstyle{1\over 4}}}
\def\eigh{{\textstyle{1\over 8}}}

\def\lsim{\mathrel{\rlap{\lower4pt\hbox{\hskip1pt$\sim$}}
    \raise1pt\hbox{$<$}}}
\def\gsim{\mathrel{\rlap{\lower4pt\hbox{\hskip1pt$\sim$}}
    \raise1pt\hbox{$>$}}}
\def\sqr#1#2{{\vcenter{\vbox{\hrule height.#2pt
         \hbox{\vrule width.#2pt height#1pt \kern#1pt
         \vrule width.#2pt}
         \hrule height.#2pt}}}}

\def\prt{\partial}

\def\etal{{\it et al.}}

\def\ol#1{\overline{#1}}

\newcommand{\beq}{\begin{equation}}
\newcommand{\eeq}{\end{equation}}
\newcommand{\bea}{\begin{eqnarray}}
\newcommand{\eea}{\end{eqnarray}}
\newcommand{\bit}{\begin{itemize}}
\newcommand{\eit}{\end{itemize}}
\newcommand{\rf}[1]{(\ref{#1})}

\def\re{{\rm Re}~}
\def\im{{\rm Im}~}

\def\mn{{\mu\nu}}

\def\kl{{\ka\la}}

\def\rs{{\rh\si}}

\def\q{q}
\def\qHat{\widehat{\q}{}}
\def\qd#1{{\q}^{(#1)}}

\def\s{s}
\def\sHat{\widehat{\s}{}}
\def\sd#1{{\s}^{(#1)}}

\def\st{\ol{\s}}

\def\std#1{{\st}^{~(#1)}}

\def\k{k}
\def\kHat{\widehat{\k}{}}
\def\kd#1{{\k}^{(#1)}}

\def\hb{\ol h{}}
\def\Mb{\ol M{}}
\def\Mbb{\ol \Mb{}}
\def\dc{\circ}

\def\mbf#1{\mbox{\boldmath$#1$}}
\def\pvec{\mbf p}
\def\phat{\mbf{\hat p}}
\def\navec{\mbf\nabla}
\def\rvec{\mbf r}
\def\rhat{\mbf{\hat r}}

\def\xvec{\mbf x}

\def\xthat{\mbf{\hat x}_3}

\def\kjm#1#2#3{k^{(#1)}_{(#2)#3}}

\def\kIdjm#1#2{\kjm{#1}{I}{#2}}
\def\kEdjm#1#2{\kjm{#1}{E}{#2}}
\def\kBdjm#1#2{\kjm{#1}{B}{#2}}
\def\kVdjm#1#2{\kjm{#1}{V}{#2}}

\def\KNdjm#1#2{{\mathcal K}^{(#1)}_{#2}}
\def\KNappdjm#1#2{{\mathcal K}^{(#1)\text{app}}_{#2}}

\def\kNdjm#1#2{k^{{\rm N}(#1)}_{#2}}
\def\kNappdjm#1#2{k^{{\rm N}(#1){\rm app}}_{#2}}
\def\kNlabdjm#1#2{k^{{\rm N}(#1){\rm lab}}_{#2}}

\def\keff#1{(\bar k_{\rm eff})_{#1}}

\begin{document}

\begin{frontmatter}

\title{
Testing local Lorentz invariance with short-range gravity}

\author{V.\ Alan Kosteleck\'y$^1$ and Matthew Mewes$^2$}

\address{$^1$Physics Department, Indiana University,
Bloomington, Indiana 47405, USA\\
$^2$Physics Department, California Polytechnic State University, 
San Luis Obispo, California 93407, USA}

\date{IUHET 622, November 2016}
\address{}
\address{\rm 
\preprintdate}

\begin{abstract}

The Newton limit of gravity is studied 
in the presence of Lorentz-violating gravitational operators 
of arbitrary mass dimension.
The linearized modified Einstein equations are obtained 
and the perturbative solutions are constructed and characterized.
We develop a formalism for data analysis 
in laboratory experiments testing gravity at short range
and demonstrate that these tests 
provide unique sensitivity to deviations from local Lorentz invariance.

\end{abstract}

\end{frontmatter}

General relativity (GR) is founded 
on the Einstein equivalence principle,
which incorporates local Lorentz invariance,
local position invariance, and the weak equivalence principle.
GR is known to provide an excellent description of classical gravity
over a broad range of length scales.
However,
modifications of the Einstein equivalence principle
associated with local Lorentz violation 
may arise in an underlying framework compatible with quantum physics
such as string theory
\cite{ksp}.
Searches for Lorentz violation in gravitational experiments
may thus yield clues about the nature of physics beyond GR
\cite{tables,lvgrreview}.

An important class of precision tests of gravity
involves experiments testing its properties 
at short distances below about a millimeter
\cite{review}.
Remarkably,
even some aspects of the conventional Newton force 
await verification on this scale,
and the presence of larger forces 
falling as an inverse cubic, quartic, or faster 
is still compatible with existing experimental data.
In this work,
we use a comprehensive description 
of possible deviations from local Lorentz invariance 
in the pure-gravity sector 
to study laboratory tests of gravity at short range 
and to characterize their sensitivity
{\it vis-\`a-vis} other types of investigations.
Our results also provide a formalism for the analysis of data
in short-range experiments.

One approach to studying Lorentz violation in gravity
is to build a specific model and study its properties.
However,
since no compelling signals for Lorentz violation
have been uncovered to date,
guidance for a broad-based experimental search
is perhaps best obtained by developing instead 
a framework allowing all types of Lorentz violation
while including accepted gravitational physics.
Effective field theory is
one powerful technique along these lines,
as it permits a general description of emergent effects 
from an unobservable scale
\cite{eft}.

In the context of gravity,
the effective field theory for Lorentz violation 
\cite{akgrav}
offers a model-independent framework
for exploring observables for Lorentz violation.
In the pure-gravity sector in Riemann geometry,
the action of this theory
contains the Einstein-Hilbert action and a cosmological constant
along with all coordinate-independent terms 
involving gravitational-field operators.
The pure-gravity action
is a subset of the general effective field theory
describing matter and gravity
known as the gravitational Standard-Model Extension (SME).
A term violating Lorentz invariance in the action
consist of a Lorentz-violating operator
contracted with a coefficient for Lorentz violation
that controls the magnitude of the resulting physical effects.
It is often convenient to classify the operators
according to their mass dimension $d$ in natural units,
with operators having larger $d$ 
likely to induce smaller physical effects at low energies
due to a greater suppression by powers of the Newton gravitational constant
or,
equivalently,
by inverse powers of the Planck mass.

To date,
comparatively few of the coefficients for Lorentz violation 
in the pure-gravity sector have been constrained
\cite{tables}.
Most remain unexplored,
and some could even involve large Lorentz violation 
that has escaped detection so far due to ``countershading'' by feeble couplings
\cite{kt09}.
For $d=4$,
certain Lorentz-violating operators 
generate noncentral orientation-dependent corrections 
to the inverse-square law.
These have been the subject of both theoretical work
\cite{bk,se09,al10,kt11,bt11,jt12,yb15,je15}
and observation
\cite{2007Battat,2007MullerInterf,2009Chung,%
2010Bennett,2012Iorio,2013Bailey,2014Shao,he15,le16,bo16}
and two-sided constraints at various levels 
down to parts in $10^{11}$ have been obtained
on the nine corresponding coefficients for Lorentz violation.
At $d=6$, 
many Lorentz-violating operators produce instead 
corrections to Newton's law involving an inverse {\it quartic} force
\cite{bkx}.
A variety of short-range experiments 
\cite{lk,hust15,hustiu}
have attained sensitivities of order $10^{-9}$ m$^{2}$ 
to the 14 combinations of pure-gravity coefficients 
controlling this type of Lorentz violation in the nonrelativistic limit,
and there are excellent prospects for improved sensitivity 
\cite{hust16}.
Constraints on some operators of dimensions $d\leq 10$
have also been reported,
based on the nonobservation of gravitational \v Cerenkov radiation 
\cite{kt15,jt16}
and from data on gravitational waves
\cite{km16},
while proposals for other measurements exist
\cite{qb16,tz16,yu16,ms16}.

To provide a comprehensive dicussion 
of possible effects of Lorentz violation in the nonrelativistic limit
relevant for short-range tests of gravity,
we can expand the metric $g_{\mn}$ 
around the Minkowski spacetime metric $\et_{\mn}$
and work with the general gauge-invariant 
and Lorentz-violating Lagrange density $\cL$,
restricting attention to terms 
quadratic in the dimensionless metric fluctuation 
$h_{\mn} \equiv g_{\mn} - \et_{\mn}$
and neglecting the cosmological constant.
In this limit,
the Einstein-Hilbert term takes the form 
\beq
\cL_{0} = 
\quar \ep^{\mu\rh\al\ka} \ep^{\nu\si\be\la} 
\et_\kl h_\mn \prt_\al\prt_\be h_\rs.
\eeq
Incorporating both Lorentz-violating and Lorentz-invariant operators 
of arbitrary mass dimension $d$,
the Lagrange density $\cL$ can be written as
\cite{km16}
\beq
\cL = \cL_0 + 
\quar h_\mn 
(\sHat{}^{\mu\rh\nu\si} 
+ \qHat{}^{\mu\rh\nu\si} 
+ \kHat{}^{\mu\nu\rh\si})
h_\rs.
\label{lag}
\eeq
Here,
the derivative operators
$\sHat{}^{\mu\rh\nu\si}$,
$\qHat{}^{\mu\rh\nu\si}$,
and $\kHat{}^{\mu\rh\nu\si}$
can be expanded as sums 
of constant cartesian coefficients
$\sd{d}{}^{\mu_1\ldots\mu_{d+2}}$,
$\qd{d}{}^{\mu_1\ldots\mu_{d+2}}$,
$\kd{d}{}^{\mu_1\ldots\mu_{d+2}}$ 
for Lorentz violation
contracted with factors of derivatives $\prt_\mu$,
\bea
&&
\sHat{}^{\mu\rh\nu\si} = 
\sum_{d\geq 4, {\rm ~even}} \sd{d}{}^{\mu\rh\dc\nu\si\dc^{d-3}} ,
\notag \\
&&
\qHat{}^{\mu\rh\nu\si} = 
\sum_{d\geq 5, {\rm ~odd}} \qd{d}{}^{\mu\rh\dc\nu\dc\si\dc^{d-4}},
\notag \\
&&
\kHat{}^{\mu\nu\rh\si} = 
\sum_{d\geq 6, {\rm ~even}} \kd{d}{}^{\mu\dc\nu\dc\rh\dc\si\dc^{d-5}} ,
\label{coeffs}
\eea
where a circle index $\dc$ denotes
an index contracted into a derivative,
and where $n$-fold contractions are written as $\dc^n$.
The operator 
$\sHat{}^{\mu\rh\nu\si}$
is antisymmetric in both the first and second pairs of indices,
while $\qHat{}^{\mu\rh\nu\si}$
is antisymmetric in the first pair 
and symmetric in the second,
and $\kHat{}^{\mu\nu\rh\si}$ is totally symmetric.
Contracting any one of these operators with a derivative produces zero. 
Note that the $d=4$ piece of $\sHat{}^{\mu\rh\nu\si}$
includes a term of the same form as $\cL_0$
with an overall scaling factor,
which can be set to zero if desired.

In studying the nonrelativistic limit,
it is convenient to work 
with the trace-reversed metric fluctuation
\beq
\hb_\mn = r_{\mn}{}^{\rs} h_{\rs} ,
\eeq
where
\beq
r_{\mn}{}^{\rs} =
\half(\et_\mu{}^\rh \et_\nu{}^\si + \et_\mu{}^\si \et_\nu{}^\rh
- \et_{\mn} \et^{\rs}) 
\eeq
is the trace-reverse operator.
The modified linearized Einstein tensor obtained by the variation of $\cL$
can be written as the sum of the usual linearized Einstein tensor $G_L^\mn$ 
and a correction $\de G_L^\mn$, 
\bea
\hskip -10pt
G_L^\mn + \de G_L^\mn
&=& \half\big(
\prt_\rh \prt^{(\mu} \hb^{\nu)\rh}
-\et^\mn \prt_\rh\prt_\si \hb^\rs - \prt^2 \hb^\mn
\big)
+ \de G_L^\mn
\notag \\
&=& - \half \prt^2 \hb^\mn
+ \de G_L^\mn,
\eea
where in the last line we adopt the Hilbert gauge, 
$\prt_\mu \hb^\mn = 0$.
The correction $\de G_L^\mn$ can be expressed as the action
of a combination of derivative operators on $\hb^\mn$,
\beq
\de G_L^\mn = \de\Mb^{\mn\rs}\hb_\rs,
\eeq
where
\beq
\de\Mb^{\mn\rs} = \de M{}^{\mn\kl} r_{\kl}{}^{\rs} 
\eeq
with
\bea
\de M{}^{\mn\rs} &=&
-\quar( \sHat{}^{\mu\rh\nu\si} +\sHat{}^{\mu\si\nu\rh} ) 
-\half \kHat{}^{\mu\nu\rh\si}
\notag \\
&&
- \eigh ( 
\qHat{}^{\mu\rh\nu\si} +\qHat{}^{\nu\rh\mu\si}
+\qHat{}^{\mu\si\nu\rh} +\qHat{}^{\nu\si\mu\rh} ) 
\eea
being expressed in terms of the operators 
appearing in the Lagrange density \rf{lag}.

The modified linearized Einstein equation takes the form
\beq
G_L^\mn + \de G_L^\mn = 8 \pi G_N T^\mn,
\label{meeq}
\eeq
where $T^\mn$ is the energy-momentum tensor. 
The trace-reversed metric fluctuation can be expanded as 
$\hb^\mn = \hb_0^\mn + \de\hb^\mn$,
where $\hb_0^\mn$ is a conventional Lorentz-invariant solution
and $\de\hb^\mn$ is the perturbation 
arising from the correction $\de G_L^\mn$.
Solving Eq.\ \rf{meeq} at first order
then reduces to solving the coupled set of equations 
\beq
\prt^2 \hb_0^\mn = - 16 \pi G_N T^\mn ,
\quad 
\prt^2 \de\hb^\mn = 2\de\Mb^\mn{}_\rs\hb_0^\rs .
\label{pert}
\eeq
In the static limit, 
the zeroth-order solution satisfies the usual Poisson equation
$\navec^2 \hb_0^\mn = -16 \pi G_N T^\mn$
and takes the standard form
\beq
\hb_0^\mn(\xvec) =
4G_N \int d^3x'\, \frac{T^\mn(\xvec')}{|\xvec-\xvec'|},
\eeq
while the first-order solution is found to be
\beq
\de\hb^\mn =
4G_N ~\de\Mb^\mn{}_\rs
\int d^3x'\, |\xvec-\xvec'|\, T^\rs(\xvec') .
\eeq
Note that this solution is compatible 
with the Hilbert gauge because 
$\prt_\mu \de\Mb^\mn{}_\rs = 0$.

For applications to short-range experiments,
which involve nonrelativistic sources,
$T^\mn$ is well approximated 
by its energy-density component $T^{00} = \rh(\xvec)$,
where $\rh(\xvec)$ is the local mass density.
We disregard here possible Lorentz-violating modifications
to the dispersion relations for various SME matter species
\cite{enmomlv,kt11},
which generate geodesics on Finsler spacetimes
\cite{finsler,finsler2}.
Also,
the components of the metric fluctuation 
can be expressed in terms of
a modified gravitational potential $U(\xvec)$
producing a modified gravitational acceleration 
$\mbf g(\xvec) = \navec U$,
\beq
h^{00} = \half \hb^{00} = 2U , 
\quad
h^{jk} = \half \hb^{00} \de^{jk} = 2U\de^{jk} .
\eeq
Expanding $U(\xvec)=U_0(\xvec)+\de U (\xvec)$ 
as the sum of the usual gravitational potential $U_0$
and the perturbation $\de U$ then yields
\beq
U_0(\xvec) = G_N \int d^3x'\, \frac{\rh(\xvec')}{|\xvec-\xvec'|},
\eeq
as expected.
The Lorentz-violating modification to the potential is given by
\bea
\de U(\xvec)
& \hskip -5pt = & \hskip -5pt
\half \de h_{00} = \half r_{00\mn} \de\hb^\mn
\notag \\
& \hskip -5pt = & \hskip -5pt
2G_N \de\Mbb_{0000} \int d^3x'\, |\xvec-\xvec'|\, \rh(\xvec') ,
\label{soln}
\eea
where for convenience we define the double trace-reversed operator
\bea
\de\Mbb_{0000}
&=& r_{00\mn} r_{00\rs} \de M^{\mn\rs} 
= \quar \de M^{\rh\rh\si\si} 
\notag\\
&=& -\eigh
( \sHat^{\rh\si\rh\si} + \kHat^{\rh\rh\si\si} ).
\label{moooo}
\eea
Note the noncovariant traces.

The last expression in Eq.\ \rf{moooo}
reveals that terms in $\cL$ 
involving the CPT-odd operator $\qHat{}^{\mu\rh\nu\si}$ 
produce no effects on short-range experiments.
Modifications of the potential in this limit
therefore arise only from operators of even dimension $d$,
which contain an even number $d-2$ of derivatives $\navec$.
Note also that the expression \rf{soln}
for the Lorentz-violating potential 
holds in regions sufficiently far from source masses
for the perturbative approach to be valid.

To make further progress in characterizing the result \rf{soln},
it is convenient to perform Fourier transforms
and work in momentum space,
where we can identify $\prt_\mu \to i p_\mu$.
In the quasistatic nonrelativistic limit
we can neglect the frequencies $p_0$,
so each derivative contraction can be taken
as a contraction with the three-momentum $p^j$.
In this limit, 
the operators $\sHat^{\rh\si\rh\si}$ and $\kHat^{\rh\rh\si\si}$
in Eq.\ \rf{moooo}
reduce to $d-2$ symmetrized three-momenta
contracted with constant coefficients for Lorentz violation.
We can therefore perform a spherical-harmonic expansion,
\beq
\de\Mbb_{0000} = \sum_{djm}
p^{d-2} Y_{jm}(\phat) \KNdjm{d}{jm} ,
\label{se}
\eeq
which captures the rotational properties of the perturbation
that are essential for short-range experiments.
Here,
the sum ranges over $d=4,6,8,\ldots$ and $j = 0,2,4,\ldots,d-2$,
and we write $p = |\pvec|$ for later convenience.
The spherical coefficients $\KNdjm{d}{jm}$ 
are constants controlling the magnitudes of the perturbative effects,
and they are linear combinations of the cartesian coefficients
appearing in Eq.\ \rf{coeffs}.

When applied to the perturbed potential \rf{soln},
the expansion \rf{se} offers some direct insights
into the nature of the perturbative effects.
The maximum angular momentum $j_{\rm max} = d-2$
arises from the totally traceless and symmetric combination 
of cartesian coefficients contained in the expressions \rf{coeffs},
which contain no Laplace operator $\navec^2$.
The contributions with $j = j_{\rm max}-2 = d-4$
arise from coefficient combinations involving one spatial trace
and therefore only a single Laplace operator.
All other terms in the expansion \rf{se} 
involve two or more spatial traces,
producing two or more Laplace operators,
and these cannot correct the gravitational potential
because $\navec^4 |\xvec-\xvec'| = 0$ outside the source.
In short,
for fixed $d$ 
$\de U$ acquires contributions
only for $j=d-2$ and $j=d-4$,
involving a total of $4d-10$ independent coefficients $\KNdjm{d}{jm}$.
This implies that for fixed $d$
only $4d-10$ independent physical effects
can modify the Newton potential
up to an overall scaling factor.
For $d=4$,
this reproduces the degrees of freedom
found in the modified potential in Eq.\ (137) of Ref.\ \cite{bk},
while for $d=6$ it matches the counting 
obtained in Eq.\ (7) of Ref.\ \cite{bkx} 
when the Lorentz-invariant trace is removed.

To gain further insight,
we can calculate $\de U(\xvec)$ for a point source,
working in momentum space for convenience.
This involves the Fourier transform of the modulus $r = |\xvec-\xvec'|$ 
of the displacement vector $\rvec = \xvec-\xvec'$
from the source mass at $\xvec'$ to the point $\xvec$,
\beq
r = |\rvec| = |\xvec-\xvec'|
= \int d^3p\ \tilde r(\pvec) e^{i\pvec\cdot\rvec} .
\eeq
This expression contains an infrared divergence
because $r$ grows at infinity.
Moreover,
the momentum-space expression for $\de U(\xvec)$ 
involves $d-2$ derivatives of $r$,
which introduces ultraviolet divergences as well.
We can control all the divergences by introducing
a regulated version of the Fourier transform $\tilde r(\pvec)$,
given by
\beq
\tilde r(\pvec;\ep,\La) =
-\frac{\prt}{\prt\ep} \frac{\ep f(\tfrac{p}{\La})}{\pi^2(p^2+\ep^2)^2} ,
\eeq
where $\ep$ regulates the infrared divergence
and $\La$ regulates the ultraviolet divergences.
The function $f(x)$ is taken as a generic even smoothing function of $x$
that is assumed to obey
$f(x)\to 1$ when $x\to 0$
and $f(x)\to 0$ when $x\to \pm \infty$
and that vanishes sufficiently rapidly 
to suppress any relevant divergences.
In imposing the limiting condition for $x\to \pm \infty$,
we are allowing for an extension of the range of 
$p = |\pvec|\geq 0$ to the full real line
for later convenience.
The physical result can ultimately be obtained by taking the limits
$\ep\rightarrow 0$ and $\La\rightarrow\infty$.
Adopting this regularization,
the perturbation of the gravitational potential
due to a single point source mass $m_s$ 
becomes
\bea
\de U(\rvec) 
& \hskip -5pt = & \hskip -5pt
- 2 G_N m_s \frac{\prt}{\prt\ep} \sum_{djm} \KNdjm{d}{jm}
\notag\\
&&
\times \int 
d\ph\, du\, dp ~
p^d Y_{jm}(\phat) 
\frac{\ep f(\tfrac{p}{\La})}{\pi^2(p^2+\ep^2)^2} e^{ipru}, 
\eea
where $u = \rhat\cdot\phat$.
Since $d\geq 4$,
the infrared divergence is no longer an issue,
so at this point we can take the limit $\ep\rightarrow 0$.

To perform the integral explicitly,
it is useful to work in a chosen ``apparatus'' frame 
in which the $x_3$ axis points along a symmetry axis of the system.
Note that this frame typically differs from
the canonical laboratory frame used in Lorentz-violation studies
\cite{sunframe}.
Here,
we adopt an apparatus frame in which $\rvec =r\xthat$
is aligned with the $x_3$ axis.
The spherical harmonics then contribute only for $m=0$
and so reduce to Legendre polynomials,
giving
\bea
\hskip -10pt
\de U^{\rm app}(r\xthat)
& \hskip -5pt = & \hskip -5pt
-\frac{2 G_N m_s}{\pi} \sum_{dj} \KNappdjm{d}{j0}
\sqrt{\tfrac{2j+1}{4\pi}}
(ir)^{4-d}
\notag\\
&&
\hskip -10pt
\times
\int_{-1}^1 du\,
P_j(u)
\left(\frac{\prt}{\prt u} \right)^{d-4}
\int_{-\infty}^\infty dp\,
f(\tfrac{p}{\La})
e^{ipru} ,
\eea
where we have used the evenness of $d$ and $j$ 
to extend the $p$ integral over the entire real line.
In the limit $\La\rightarrow\infty$, 
the $p$ integral becomes $2\pi\de(ru)$.
Performing the $u$ integral using this delta function yields 
\bea
\hskip-10pt
\de U^{\rm app}(r\xthat) 
& \hskip -5pt = & \hskip -5pt
- 4 G_N m_s
\sum_{dj} \KNappdjm{d}{j0}
\notag\\
&&
\hskip 40pt
\times 
(-1)^{d/2}
\sqrt{\tfrac{2j+1}{4\pi}}
r^{3-d}     
P^{(d-4)}_j(0) ,
\eea
where $P^{(n)}_j(x)$ denotes the $n$-th derivative
of the Legendre polynomial $P_j(x)$.
Notice that this result is zero unless either $j=d-2$ or $j=d-4$,
as expected,
because the Legendre polynomial $P_j$ is of order $j$.
Evaluation of the $(d-4)$ derivatives of the Legendre polynomial gives
\beq
P^{(d-4)}_j(0) =
\frac{(-1)^{(d+j)/2}(j+d-4)!}
{{2^j}\left[\half (j-d+4)\right]! \left[\half (j+d-4)\right]!},
\eeq
showing that they are nonvanishing for the range of indices of interest here.
The correction to the Newton gravitational potential
in the apparatus frame
containing all contributing effects for Lorentz violation 
can therefore be written in the compact form
\beq
\de U^{\rm app}(r\xthat) = 
\sum_{dj}
\frac{G_N m_s}{r^{d-3}}
\sqrt{\tfrac{2j+1}{4\pi}}
\kNappdjm{d}{j0} ,
\label{app}
\eeq
where now the sum over $d$ includes 
even values $d\geq 4$,
and the allowed values of $j$ are $j=d-2$ and $d-4$.
In this equation,
we have introduced reduced spherical coefficients for
Lorentz violation
defined as 
\beq
\kNdjm{d}{jm} \equiv 4 (-1)^{1+d/2} P^{(d-4)}_j(0) ~\KNdjm{d}{jm} .
\eeq
Note that this definition holds in any frame.

To apply this result in realistic circumstances,
we must reconstruct the gravitational potential 
in the canonical laboratory frame
\cite{sunframe}.
In this frame,
the $z$ axis points towards the zenith
and the $x$ axis lies at an angle $\vp$ east of south.
Using standard angles for spherical polar coordinates, 
we can write the components of the displacement vector as
$\rvec = r (\cos\ph\sin\th, \sin\ph\sin\th, \cos\th)$.
The linear combination 
of spherical coefficients in the laboratory frame 
producing the spherical coefficients in the apparatus frame 
involves a rotation and can be expressed as
\beq
\kNappdjm{d}{jm} 
= \sum_{m'} e^{im\ga} e^{im'\ph} d^{(j)}_{mm'}(-\th) \kNlabdjm{d}{jm'} ,
\eeq
where $\ga$ is an Euler angle relating the two frames
that plays no physical role
because the sum \rf{app} involves only $m=0$,
and where the quantities $d^{(j)}_{mm'}(\be)$
are the little Wigner matrices
given by Eq.\ (136) of Ref.\ \cite{km09}.
Using the identity
\beq
Y_{jm}(\th,\ph) =
\sqrt{\tfrac{2j+1}{4\pi}}
e^{im\ph} d^{(j)}_{0m}(-\th) ,
\eeq
we find that the correction to the gravitational potential
in terms of the spherical coordinates $r$, $\th$, and $\ph$
in the laboratory frame is
\beq
\de U(\rvec)
= \sum_{djm}
\frac{G_N m_s}{r^{d-3}} Y_{jm}(\th,\ph) \kNlabdjm{d}{jm} ,
\label{labpot}
\eeq
where $d\geq 4$ is even,
$j=d-2$ or $j=d-4$,
and $m=-j,\ldots j$.
The corresponding correction $\de \mbf g(\rvec)$
to the gravitational acceleration is given by
\beq
\de \mbf g (\rvec) = \navec \de U.
\eeq

The coefficients $\kNdjm{d}{jm}$ for Lorentz violation
are frame-dependent quantities,
so an inertial frame must be specified in reporting their measurement.
The laboratory frame is noninertial
due to the rotation of the Earth,
so it is unsuitable for this purpose.
The canonical inertial frame adopted in the literature
is the Sun-centered celestial-equatorial frame 
\cite{sunframe},
which is conventionally defined using cartesian coordinates $(T,X,Y,Z)$.
The origin for $T$ is fixed as the 2000 vernal equinox,
at which time the $X$ axis 
lies along the line from the Earth to the Sun.
The $Z$ axis is aligned with the rotation axis of the Earth,
and the $Y$ axis forms a right-handed coordinate system.
To a sufficient approximation,
the Sun-centered frame is inertial 
over the time scale of typical laboratory experiments,
and in this frame the coefficients for Lorentz violation
can be taken as spacetime constants 
\cite{ck}. 
The Earth rotation therefore induces variations with sidereal time 
of the coefficients in the laboratory frame,
which implies that sidereal variations can appear in experimental data 
\cite{ak98}.
The result \rf{labpot} must therefore be expressed
in terms of coefficients in the Sun-centered frame
when performing an experimental analysis.
This conversion involves a rotation that depends on sidereal time
\cite{sunframe}.
Standard methods 
\cite{km09}
can be applied to obtain the relationship 
\beq
\kNlabdjm{d}{jm}
= \sum_{m'} e^{im\varphi}e^{im'\om_\oplus T_\oplus} 
d^{(j)}_{mm'}(-\ch)\,
\kNdjm{d}{jm'} 
\label{scf}
\eeq
between the Newton spherical coefficients $\kNlabdjm{d}{jm}$ 
in the laboratory frame and 
the Newton spherical coefficients $\kNdjm{d}{jm}$ in the Sun-centered frame.
Here,
$\om_\oplus \simeq 2\pi/({\rm 23~h~56~m})$
is the sidereal angular rotation rate of the Earth,
and $\ch$ is the colatitude of the laboratory.
Also,
$T_\oplus$ is the local laboratory sidereal time,
which differs from the time $T$ by a constant offset 
\cite{offset}.

The coefficients $\kNdjm{d}{jm}$ in the canonical Sun-centered frame
are the ultimate target of experimental analyses. 
Using the result \rf{labpot}
from a point source of mass $m_s$
together with spherical coefficients expressed in the Sun-centered frame
according to Eq.\ \rf{scf},
the gravitational potential and hence the force
due to an extended source mass can be obtained.
The inverse-power corrections appearing in the potential \rf{labpot}
imply that experiments testing gravity at short range 
have maximal sensitivity to these Lorentz-violating effects.
In practical applications,
numerical methods are likely to be required
to calculate the gravitational potential from a test mass of finite extent
\cite{lk,hust15,hustiu,hust16}.
Nonetheless,
the equations derived here via the spherical decomposition 
provide a clean separation
of the observable harmonics in sidereal time
and therefore offer a direct path
for analyses seeking effects of Lorentz violation at arbitrary $d$.

The methodology developed here also permits sensitivity comparisons 
between short-range experiments and other types of investigations.
For example,
an earlier analysis has provided a complete characterization
of coefficients for Lorentz violation 
that are accessible to experiments 
involving gravitational waves
\cite{km16}.
This work reveals that Lorentz violation in gravitational radiation 
is controlled by four sets of vacuum spherical coefficients,
$\kIdjm{d}{jm}$,
$\kEdjm{d}{jm}$,
$\kBdjm{d}{jm}$,
and $\kVdjm{d}{jm}$.
Data from the observation of gravitational waves
and the absence of gravitational \v Cerenkov radiation 
already place significant constraints on
a subset of these coefficients
\cite{kt15,km16}.
It is therefore of definite interest
to establish the relationship
between the Newton spherical coefficients $\kNdjm{d}{jm}$ 
and the vacuum spherical coefficients
in the canonical Sun-centered frame.
Here,
we demonstrate that a partial overlap exists:
short-range experiments are sensitive
to certain types of Lorentz violation
that are inaccessible to analyses using gravitational radiation,
and vice versa.

\renewcommand{\arraystretch}{1.4}
\begin{table*}
\begin{center}
\begin{tabular}{c|c||c|c|c|c|c|c|c|c|c}
Type	&	Coefficient	&	Parity	&	$d$	&	$j$	&	Number	&	$d=4$	&	$d=5$	&	$d=6$	&	$d=7$	&	$d=8$	\\
		\hline\hline																			
Newton	&	$\kNdjm{d}{jm}$	&	$E$	&	even, $\geq4$	&	$d-4,d-2$	&	$4d-10$	&	6	&	--	&	14	&	--	&	22	\\
		\hline																			
vacuum	&	$\kIdjm{d}{jm}$	&	$E$	&	even, $\geq4$	&	$0,1,\ldots,d-2$	&	$(d-1)^2$   	&	9	&	--	&	25	&	--	&	49	\\
	&	$\kEdjm{d}{jm}$	&	$E$	&	even, $\geq6$	&	$4,5,\ldots,d-2$	&	$(d-1)^2-16$	&	--	&	--	&	9	&	--	&	33	\\
	&	$\kBdjm{d}{jm}$	&	$B$	&	even, $\geq6$	&	$4,5,\ldots,d-2$	&	$(d-1)^2-16$	&	--	&	--	&	9	&	--	&	33	\\
	&	$\kVdjm{d}{jm}$	&	$B$	&	odd,  $\geq5$	&	$0,1,\ldots,d-2$	&	$(d-1)^2$	&	--	&	16	&	--	&	36	&	--	\\
\end{tabular}
\caption{\label{sphcoeffs}
Summary of Newton and vacuum spherical coefficients.}
\end{center}
\end{table*}

Table \ref{sphcoeffs} provides a summary
of the Newton and vacuum spherical coefficients.
The first two columns of this table identify the type of coefficients
and list their components.
The third column lists the parity of the corresponding operator. 
The next two columns give
the allowed ranges of $d$ and $j$.
The sixth column lists the number of independent components
of the coefficients.
The remaining columns provide explicit values 
of the number of independent components for $4\leq d \leq 8$
for convenience.
We remark in passing that 
the vacuum spherical coefficients $\kIdjm{d}{jm}$
can be expressed in terms of the spherical coefficients $\ol s^{(d)}_{jm}$
introduced in Ref.\ \cite{kt15} 
in the context of studies of gravitational \v Cerenkov radiation
according to
\beq
\kIdjm{d}{jm} = 
\half (-1)^{1+j+d/2} 
\ol s^{(d)}_{jm} ,
\label{ktcoeffs}
\eeq
which is a one-to-one relationship.

For the case $d=4$,
the only coefficients for Lorentz violation
appearing in the Lagrange density \rf{lag}
are the cartesian coefficients $\sd{4}{}^{\mu\rh\al\nu\si\be}$.
These contain ten independent components,
which can conveniently be packaged in the dual cartesian coefficients
\beq
\std{4}_{\kl} 
\equiv 
-\tfrac 1{36}\ep_{\mu\rh\al\ka}\ep_{\nu\si\be\la}
\sd{4}{}^{\mu\rh\al\nu\si\be}.
\eeq
Note that the cartesian indices distinguish these dual coefficients
from the related spherical coefficients appearing in Eq.\ \rf{ktcoeffs}.
Calculation shows that 
the nine vacuum spherical coefficients $\kIdjm{d}{jm}$
are given in terms of the dual cartesian coefficients $\std{4}_{\kl}$ by
\bea
\kIdjm{4}{00}
& \hskip -5pt = & \hskip -5pt 
-\sqrt{\tfrac{\pi}{9}}~
( \std{4}_\mn \et^\mn + 4\std{4}_{tt}),
\notag\\
\kIdjm{4}{10}
& \hskip -5pt = & \hskip -5pt 
\sqrt{\tfrac{4\pi}{3}} 
~\std{4}_{tz},
\notag\\
\re\kIdjm{4}{11}
& \hskip -5pt = & \hskip -5pt 
-\sqrt{\tfrac{2\pi}{3}} 
~\std{4}_{tx},
\notag\\
\im\kIdjm{4}{11}
& \hskip -5pt = & \hskip -5pt 
\sqrt{\tfrac{2\pi}{3}} 
~\std{4}_{ty},
\notag\\
\kIdjm{4}{20}
& \hskip -5pt = & \hskip -5pt 
-\sqrt{\tfrac{\pi}{5}}~
(\tfrac {2}{3} \std{4}_\mn \et^\mn 
+\tfrac {2}{3} \std{4}_{tt}
- \std{4}_{xx} - \std{4}_{yy}),
\notag\\
\re\kIdjm{4}{21}
& \hskip -5pt = & \hskip -5pt 
\sqrt{\tfrac{4\pi}{30}} 
~\std{4}_{xz},
\notag\\
\im\kIdjm{4}{21}
& \hskip -5pt = & \hskip -5pt 
-\sqrt{\tfrac{4\pi}{30}} 
~\std{4}_{yz},
\notag\\
\re\kIdjm{4}{22}
& \hskip -5pt = & \hskip -5pt 
-\sqrt{\tfrac{\pi}{30}} 
~(\std{4}_{xx} -\std{4}_{yy}),
\notag\\
\im\kIdjm{4}{22}
& \hskip -5pt = & \hskip -5pt 
\sqrt{\tfrac{4\pi}{30}} 
~\std{4}_{xy}.
\eea
We can also show that the six Newton spherical coefficients $\kNdjm{4}{jm}$
are related to the dual cartesian coefficients $\std{4}_{\kl}$
according to
\bea
\kNdjm{4}{00}
& \hskip -5pt = & \hskip -5pt 
\sqrt{\tfrac{4\pi}{9}}~
( 2\std{4}_\mn \et^\mn + 5\std{4}_{tt}),
\notag\\
\kNdjm{4}{2m}
& \hskip -5pt = & \hskip -5pt 
-\kIdjm{4}{2m},
\eea
which reveals that 
the two sets of $j=2$ spherical coefficients are equal.
Indeed,
the nine vacuum spherical coefficients $\kIdjm{4}{jm}$
and the isotropic Newton coefficient $\kNdjm{4}{00}$ 
together span the ten-dimensional coefficient space for $d=4$.
Moreover,
the above equations involve also the trace $\std{4}_\mn \et^\mn$,
which governs Lorentz-invariant effects.
Setting this to zero reduces the total number of independent 
cartesian coefficients to nine and implies 
\beq
\kNdjm{4}{00} = -\tfrac 52 \kIdjm{4}{00}
= \tfrac{10\sqrt\pi}{3}~\std{4}_{tt} .
\label{newtfour}
\eeq
The nine vacuum spherical coefficients $\kIdjm{4}{jm}$
therefore span the coefficient space for $d=4$,
and so the six Newton spherical coefficients $\kNdjm{4}{jm}$ 
are completely determined by the vacuum spherical coefficients 
$\kIdjm{4}{jm}$.
Since tight two-sided bounds on $\kIdjm{4}{2m}$
have been obtained from the absence of gravitational \v Cerenkov radiation
\cite{kt15},
we can conclude that short-range tests of gravity
searching for anisotropic effects
cannot yield unique information about the $d=4$ coefficients.

The situation for $d=6$ is more involved.
Here,
there are 84 independent cartesian coefficients
$\sd{6}{}^{\mu_1\ldots\mu_8}$ 
and also 105 independent cartesian coefficients
$\kd{6}{}^{\mu_1\ldots\mu_8}$, 
for a total of 189 degrees of freedom.
For $d=6$,
Table \ref{sphcoeffs} shows that 
the vacuum spherical coefficients include
25 independent components of $\kIdjm{6}{jm}$ 
governing nonbirefringent effects,
9 independent components $\kEdjm{6}{jm}$
controlling $E$-parity birefringent effects, 
and 9 independent components $\kBdjm{6}{jm}$
determining $B$-parity birefringent effects.
The Newton spherical coefficients $\kNdjm{6}{jm}$
include 14 independent cofficients 
controlling $E$-parity Lorentz-violating operators.
Explicit expressions for all 57 of these spherical coefficients
in terms of the 189 independent cartesian coefficients
are lengthy and so are omitted here.
However,
some calculation yields a relationship 
among the $j=4$ components of the spherical coefficients
controlling $E$-parity effects, 
\beq
\kNdjm{6}{4m} = 15\kIdjm{6}{4m} - \sqrt{\tfrac{45}{14}}~\kEdjm{6}{4m} .
\label{d6rel}
\eeq
This reveals that 
the nine Newton spherical coefficients $\kNdjm{6}{4m}$ with $j=4$ 
are completely determined by vacuum spherical coefficients.
In contrast,
the five Newton spherical coefficients $\kNdjm{6}{2m}$ with $j=2$ 
are independent of the vacuum spherical coefficients.

At present,
the coefficients $\kIdjm{6}{4m}$ are tightly constrained
\cite{kt15},
but the limits on the coefficients $\kEdjm{6}{4m}$
from gravitational waves 
\cite{km16}
are one to three orders of magnitude weaker 
than the best current bounds from short-range experiments
\cite{hustiu}.
This difference in sensitivity can be traced to the inverse-quartic behavior 
of the modified gravitational force
and the consequent gain in reach for tests at short range.
However,
even if future techniques for gravitational radiation
are developed that permit vastly improved sensitivities
to the coefficients $\kEdjm{6}{4m}$,
the analysis performed here demonstrates 
that the Newton spherical coefficients $\kNdjm{6}{2m}$ with $j=2$ 
will remain unconstrained by gravitational-radiation studies
while being accessible in short-range experiments.
Also,
many vacuum spherical coefficients that can be studied
using gravitational waves
are inaccessible to short-range experiments,
so the two extremes of Newton and relativistic experiments
provide a complementary sensitivity 
to violations of Lorentz invariance in the gravity sector.
Note also that the 48 independent degrees of freedom 
spanned in total by the vacuum and Newton spherical coefficients
leave a 141-dimensional coefficient space at $d=6$
that is untouched by studies of Lorentz violation
using gravitational radiation or short-range tests of gravity.
Identifying experimental tests with sensitivity 
to these many unconstrained effects
is an interesting and worthwhile open problem. 

For completeness,
we can also explicitly relate 
the 14 $d=6$ Newton spherical coefficients $\kNdjm{6}{jm}$
to the 14 $d=6$ effective cartesian coefficients $\keff{JKLM}$
adopted in the recent literature discussing searches for Lorentz violation 
with experiments on short-range gravity
\cite{bkx,lk,hust15,hustiu}.
We find the correspondence 
\bea
\kNdjm{6}{20} 
& \hskip -5pt = & \hskip -5pt 
\tfrac{36}{7}\sqrt{\tfrac{\pi}{5}}~\big(
\keff{XXJJ}+\keff{YYJJ}\big) ,
\notag\\
\re\kNdjm{6}{21} 
& \hskip -5pt = & \hskip -5pt 
\tfrac{12}{7}\sqrt{\tfrac{6\pi}{5}}\keff{XZJJ} ,
\notag\\
\im\kNdjm{6}{21} 
& \hskip -5pt = & \hskip -5pt 
-\tfrac{12}{7}\sqrt{\tfrac{6\pi}{5}}\keff{YZJJ} ,
\notag\\
\re\kNdjm{6}{22} 
& \hskip -5pt = & \hskip -5pt 
-\tfrac{6}{7}\sqrt{\tfrac{6\pi}{5}}~\big(
\keff{XXJJ}-\keff{YYJJ}\big) ,
\notag\\
\im\kNdjm{6}{22} 
& \hskip -5pt = & \hskip -5pt 
\tfrac{12}{7}\sqrt{\tfrac{6\pi}{5}}\keff{XYJJ} ,
\notag\\
\kNdjm{6}{40} 
& \hskip -5pt = & \hskip -5pt 
-\tfrac{5}{7}\sqrt{\pi} ~\big(
\keff{XXJJ}+\keff{YYJJ}
\notag\\
&&
\hskip20pt
+7\keff{XXZZ}+7\keff{YYZZ}\big) ,
\notag\\
\re\kNdjm{6}{41} 
& \hskip -5pt = & \hskip -5pt 
\tfrac{2}{7}\sqrt{5\pi} ~\big(
3\keff{XZJJ}-7\keff{XZZZ}\big) ,
\notag\\
\im\kNdjm{6}{41} 
& \hskip -5pt = & \hskip -5pt 
-\tfrac{2}{7}\sqrt{5\pi} ~\big(
3\keff{YZJJ}-7\keff{YZZZ}\big) ,
\notag\\
\re\kNdjm{6}{42} 
& \hskip -5pt = & \hskip -5pt 
-\tfrac{1}{7}\sqrt{10\pi} ~\big(
\keff{XXJJ}-\keff{YYJJ}
\notag\\
&&
\hskip20pt
-7\keff{XXZZ}+7\keff{YYZZ}\big) ,
\notag\\
\im\kNdjm{6}{42} 
& \hskip -5pt = & \hskip -5pt 
\tfrac{2}{7}\sqrt{10\pi} ~\big(
\keff{XYJJ}-7\keff{XYZZ}\big) ,
\notag\\
\re\kNdjm{6}{43} 
& \hskip -5pt = & \hskip -5pt 
-2\sqrt{\tfrac{5\pi}{7}} ~\big(
\keff{XXXZ}-3\keff{XYYZ}\big) ,
\notag\\
\im\kNdjm{6}{43}
& \hskip -5pt = & \hskip -5pt 
-2\sqrt{\tfrac{5\pi}{7}} ~\big(
\keff{YYYZ}-3\keff{XXYZ}\big) ,
\notag\\
\re\kNdjm{6}{44} 
& \hskip -5pt = & \hskip -5pt 
\sqrt{\tfrac{5\pi}{14}} ~\big(
\keff{XXXX}+\keff{YYYY}-6\keff{XXYY}\big) ,
\notag\\
\im\kNdjm{6}{44} 
& \hskip -5pt = & \hskip -5pt 
-2\sqrt{\tfrac{10\pi}{7}} ~\big(
\keff{XXXY}-\keff{XYYY}\big) ,
\label{newtsix}
\eea
where the rotation-invariant double trace $\keff{JJKK}$ is assumed zero.
Using these equations,
we can translate the independent values
of the effective cartesian coefficients $\keff{JKLM}$
provided in Table II of Ref.\ \cite{hustiu}
into constraints on the Newton spherical coefficients $\kNdjm{6}{jm}$.
Propagating the errors ignoring correlations
yields the values given in Table \ref{kefftab}.
Note that other existing results 
for cartesian coefficients with $d=4$
\cite{2007Battat,2007MullerInterf,2009Chung,%
2010Bennett,2012Iorio,2013Bailey,2014Shao,he15,le16,bo16}
and $d=6$
\cite{lk,hust15,hustiu}
can also be converted to measurements of spherical coefficients
using the correspondences \rf{newtfour} and \rf{newtsix}.
Moreover, 
certain experiments studying Lorentz-invariant short-range gravity
\cite{ka07,mu14,ho85}
and conceivably others designed to search 
for large Lorentz-invariant forces at short distances 
\cite{iupui,ge08,ge10,so14}
may have sensitivity to the Newton spherical coefficients 
$\kNdjm{d}{jm}$ through the perturbation \rf{labpot}
as well.

\renewcommand{\arraystretch}{1.4}
\begin{table}
\begin{center}
\begin{tabular}{c|c}
Coefficient & Measurement\\
\hline\hline
$	\kNdjm{6}{20}	$	&	$	(3\pm 23) \times10^{-8}\mbox{ m}^2	$	\\
$	\re\kNdjm{6}{21}	$	&	$	(-4\pm 4) \times10^{-8}\mbox{ m}^2	$	\\
$	\im\kNdjm{6}{21}	$	&	$	(-2\pm 4) \times10^{-8}\mbox{ m}^2	$	\\
$	\re\kNdjm{6}{22}	$	&	$	(0\pm 9) \times10^{-8}\mbox{ m}^2	$	\\
$	\im\kNdjm{6}{22}	$	&	$	(1\pm 4) \times10^{-8}\mbox{ m}^2	$	\\
$	\kNdjm{6}{40}	$	&	$	(4\pm 25) \times10^{-8}\mbox{ m}^2	$	\\
$	\re\kNdjm{6}{41}	$	&	$	(3\pm 5) \times10^{-8}\mbox{ m}^2	$	\\
$	\im\kNdjm{6}{41} 	$	&	$	(1\pm 5) \times10^{-8}\mbox{ m}^2	$	\\
$	\re\kNdjm{6}{42}	$	&	$	(0\pm 12) \times10^{-8}\mbox{ m}^2	$	\\
$	\im\kNdjm{6}{42}	$	&	$	(2\pm 2) \times10^{-8}\mbox{ m}^2	$	\\
$	\re\kNdjm{6}{43}	$	&	$	(0\pm 1) \times10^{-8}\mbox{ m}^2	$	\\
$	\im\kNdjm{6}{43}	$	&	$	(1\pm 1) \times10^{-8}\mbox{ m}^2	$	\\
$	\re\kNdjm{6}{44}	$	&	$	(2\pm 9) \times10^{-8}\mbox{ m}^2	$	\\
$	\im\kNdjm{6}{44}	$	&	$	(2\pm 5) \times10^{-8}\mbox{ m}^2	$	\\
\end{tabular}
\caption{\label{kefftab}
Derived values of Newton spherical coefficients.
} 
\end{center}
\end{table}

For even $d\geq 8$ the calculations are more challenging, 
but we conjecture a similar relationship
to the result \rf{d6rel},
\beq
\kNdjm{d}{d-2,m} = a_d \kIdjm{d}{d-2, m} + b_d \kEdjm{d}{d-2,m} ,
\label{drel}
\eeq
where $a_d$ and $b_d$ are real constants.
For example,
we expect that for $d=8$
the 115 independent vacuum spherical coefficients
and the 22 independent Newton spherical coefficients
can be expressed in terms of the
270+630=900 independent cartesian coefficients,
with 13 of the 22 Newton coefficients
determined in terms of vacuum spherical coefficients
according to Eq.\ \rf{drel}
and with short-range tests of gravity offering unique access 
to the remaining nine Newton spherical coefficients $\kNdjm{8}{4m}$.

To summarize,
we have developed in this work
a convenient formalism for analyzing short-range tests of gravity
for general signals of Lorentz violation.
The procedure adopts a spherical decomposition
to enable a treatment of Lorentz-violating operators
of arbitrary mass dimension
and to provide a comparatively simple description
of the predicted sidereal variations in the data.
The techniques also permit the separation of signals 
into effects that are in principle
observable via gravitational radiation
and ones that are unique to short-range tests of gravity.
The presence of the latter for all $d\geq 6$
and the exceptional sensitivity of short-range tests
to the associated inverse-power modifications 
of the gravitational potential of a point source
imply a promising future for this class of laboratory experiments.

\medskip

This work was supported in part 
by the U.S.\ Department of Energy
under grant no.\ {DE}-SC0010120,
by the U.S.\ National Science Foundation 
under grant no.\ PHY-1520570,
and by the Indiana University Center for Spacetime Symmetries.

\end{document}